\documentclass[preprint]{aastex}

\newcommand{\etal}{{et~al.}\,}      
\newcommand{\eg}{{e.g.,}\,}         

\newcommand{\Msun}{$M_{\sun}$}
\newcommand{\MSUN}{{\rm M}_\odot}

\newcommand{\yr}{\rm yr^{-1}}

\newcommand{\sfg}{S$^{4}$G}

\newcommand{\spitz}{{\it Spitzer}}
\newcommand{\HA}{\mbox{H$\alpha$}}

\newcommand{\HI}{\mbox{\normalsize H\thinspace\footnotesize I}}

\def\aspcs  {{ASP Conf. Ser. }}

\usepackage{fancyheadings}
\usepackage{amsmath}
\usepackage{amssymb}
\usepackage{graphics}
\usepackage{natbib}
\bibpunct{(}{)}{;}{a}{}{,}

\def\apj    {{ApJ }}
\def\apjl   {{ApJL }}
\def\apjs   {{ApJS }}
\def\aj     {{AJ }}

\def\mnras  {{MNRAS }}

\def\pasp   {{PASP }}

\def\aspcs  {{ASP Conf. Ser. }}

\slugcomment{to be submitted to ApJ}

\shorttitle{}
\shortauthors{de Swardt et al.}

\begin{document}


\title{The Odd Offset between the Galactic Disk and Its Bar in NGC~3906}

\author{Bonita de Swardt\altaffilmark{1,2},
Kartik Sheth\altaffilmark{3},
Taehyun Kim\altaffilmark{3},
Stephen Pardy \altaffilmark{4},
Elena D' Onghia \altaffilmark{4,5},
Eric Wilcots \altaffilmark{4},
Joannah Hinz\altaffilmark{6},
Juan-Carlos Mu\~noz-Mateos\altaffilmark{3,7},
Michael W. Regan\altaffilmark{8}
E. Athanassoula\altaffilmark{9},
Albert Bosma\altaffilmark{9},
Ronald J. Buta\altaffilmark{10}, 
Mauricio Cisternas\altaffilmark{11,12}, 
S\'ebastien Comer\'on\altaffilmark{13,14},  
Dimitri A. Gadotti\altaffilmark{7}, 
Armando Gil de Paz\altaffilmark{15}, 
Thomas H. Jarrett\altaffilmark{16}, 
Bruce G. Elmegreen\altaffilmark{17},  
Santiago Erroz-Ferrer\altaffilmark{11,12},
Luis C. Ho\altaffilmark{18,19},
Johan H. Knapen\altaffilmark{11,12}, 
Jarkko Laine\altaffilmark{13,14},  
Eija Laurikainen\altaffilmark{13,14},  
Barry F. Madore\altaffilmark{18},  
Sharon Meidt\altaffilmark{20}
Kar\'in Men\'endez-Delmestre\altaffilmark{21}
Chien Y. Peng\altaffilmark{18}, 
Heikki Salo\altaffilmark{13},  
Eva Schinnerer\altaffilmark{20}
Dennis Zaritsky\altaffilmark{22}, 
}

\altaffiltext{1}{South African Astronomical Observatory, Observatory, 7935 Cape Town, South Africa}
\altaffiltext{2}{SKA South Africa, 3rd Floor, The Park, Park Road, Pinelands, South Africa}
\altaffiltext{3}{National Radio Astronomy Observatory / NAASC, 520 Edgemont Road, Charlottesville, VA 22903, USA}
\altaffiltext{4}{Department of Astronomy, University of Wisconsin, 475 North Charter Street, Madison, WI, 53706, USA}
\altaffiltext{5}{Alfred P. Sloan Fellow}
\altaffiltext{6}{MMTO, University of Arizona, 933 North Cherry Avenue, Tucson, AZ 85721, USA}
\altaffiltext{7}{European Southern Observatory, Casilla 19001, Santiago 19, Chile}
\altaffiltext{8}{Space Telescope Science Institute, 3700 San Martin Drive, Baltimore, MD 21218, USA}
\altaffiltext{9}{Aix Marseille Universit{\'e}, CNRS, LAM (Laboratoire d'Astrophysique de Marseille) UMR 7326, 13388 Marseille, France}
\altaffiltext{10}{Department of Physics and Astronomy, University of Alabama, Box 870324, Tuscaloosa, AL 35487, USA}
\altaffiltext{11}{Instituto de Astrof\'\i sica de Canarias, E-38200 La Laguna, Tenerife, Spain}
\altaffiltext{12}{Departamento de Astrof\'\i sica, Universidad de La Laguna, E-38205 La Laguna, Tenerife, Spain}

\altaffiltext{13}{Division of Astronomy, Department of Physical Sciences, University of Oulu, Oulu, FIN-90014, Finland}
\altaffiltext{14}{Finnish Centre of Astronomy with ESO (FINCA), University of Turku, V{\"a}is{\"a}l{\"a}ntie 20, FI-21500, Piikki{\"o}, Finland}
\altaffiltext{15}{Departamento de Astrof\'\i sica, Universidad Complutense de Madrid, Madrid 28040, Spain}
\altaffiltext{16}{Astronomy Department, University of Cape Town, Rondebosch 7701, South Africa}
\altaffiltext{17}{IBM Research Division, T.J. Watson Research Center, Yorktown Hts., NY 10598, USA}
\altaffiltext{18}{The Observatories of the Carnegie Institution of Washington, 813 Santa Barbara Street, Pasadena, CA 91101, USA}
\altaffiltext{19}{Kavli Institute for Astronomy and Astrophysics, Peking University 5 Yi He Yuan Road, Haidian District, Beijing 100871, P. R. China}
\altaffiltext{20}{Max-Planck-Institut f{\"u}r Astronomie/K{\"o}nigstuhl 17 D-69117 Heidelberg, Germany}
\altaffiltext{21}{Universidade Federal do Rio de Janeiro, Observat{\'o}rio do Valongo, Ladeira Pedro Ant{\^{o}}nio, 43, CEP 20080-090, Rio de Janeiro, Brazil}
\altaffiltext{22}{University of Arizona, 933 N. Cherry Ave, Tucson, AZ 85721, USA}
\begin{abstract}

We use mid-infrared 3.6 and 4.5$\mu$m imaging of NGC~3906 from the \spitz\ Survey of Stellar Structure in Galaxies (\sfg) to understand the nature of an unusual offset between its stellar bar and the photometric center of an otherwise regular, circular outer stellar disk.  
We measure an offset of $\sim$ 720 pc between the  center of the stellar bar and photometric center of the stellar disk; the bar center coincides with the kinematic center of the disk determined from previous HI observations.  Although the undisturbed shape of the disk suggests that NGC~3906 has not undergone a significant merger event in its recent history, the most plausible explanation for the observed offset is an interaction.  Given the relatively isolated nature of NGC 3906 this interaction could be with dark matter sub structure in the galaxy's halo or from a recent interaction with a fast moving neighbor which remains to be identified.   Simulations aimed at reproducing the observed offset between the stellar bar / kinematic center of the system and the photometric  center of the disk are necessary to confirm this hypothesis and constrain the interaction history of the galaxy.
\end{abstract}

\section{Introduction}

The long unsolved problem of asymmetries in Magellanic type spirals has come under increased scrutiny due to recent kinematic studies of the Large and Small Magellanic Clouds (LMC and SMC). These studies also showed that an interaction with the SMC passing close to the center of the LMC is the most probable cause of the off centered stellar bar \citep{Bes2012}.  Off-center or "offset bars" are a characteristic feature in late-type spiral galaxies and in particular Magellanic types \citep{Vau1972, Ode1996, ElmElm80} -- the usual definition of an offset bar is a stellar bar whose photometric center is offset from the photometric center of the outer isophotes of a galaxy disk.  

Off-centered stellar bars coupled with remnant spiral structure could contribute to asymmetries in a stellar disk -- asymmetrical stellar disks are often referred to as ``lopsided'' disks.   However, a recent  study of lopsidedness in 167 galaxies  by \citet{Zar2013} from the \spitz\ Survey of Stellar Structure in Galaxies (\sfg) \citep{Sheth2010}, 
 found that lopsidedness is a generic feature of disks.   While the lopsidedness is correlated with the strength of the spiral pattern it is {\em not} correlated with the presence of a stellar bar.  \citet{Zar2013} specifically note that a stellar bar is neither the cause of lopsidedness nor is the lopsidedness giving rise to the formation of a stellar bar.   

Numerical simulations of barred galaxies have shown that a bar may become off-centered through from its disk via an interaction of a small companion \citep{Ath1996,Ath1997,Ber2003}. \citet{Bek2009} investigated possible evolutionary scenarios that show that an off-center bar with some degree of spiral structure can form in the LMC if a galaxy collides with a low-mass Galactic subhalo corresponding to only a few per cent of the total mass of the LMC. These subhaloes were  referred to as `dark satellites' since they may have no or little visible matter. \citet{Bek2009} claimed that dark satellites with masses in the range of $10^8-10^9$\Msun\ can strongly disturb the disk of low mass galaxies such as the LMC.

However there are many isolated Magellanic galaxies.  In a VLA study, \citet{Wil2004} found only two of thirteen Magellanics  were clearly  interacting with their neighbor. They  argued that in most cases the asymmetry in barred Magellanic spirals cannot be explained by on-going interactions with a companion galaxy or by environment.  This earlier result was put on solid ground with the larger \sfg\ study by \citet{Zar2013}.  So the offset bars, which are a common feature of Magellanics, may be  a long lived feature that is caused by some other process. \citet{Vau1970} suggested that the asymmetry in a disk  may suggest that it is a "protogalaxy" that might be forming a bar as well.  

NGC~3906 is an ideal case to study these odd properties of a disk and an offset bar in detail.   Classified as a Magellanic type face-on spiral, NGC~3906 hosts a displaced stellar bar from what appears to be an otherwise {\em undisturbed} circular stellar disk, showing no signature of any tidal interaction at ultra-violet and optical wavelengths.  A search with NASA's Extragalactic Database reveals that the closest companions to NGC~3906 are located 14-16$\arcmin$ ($\rm~50kpc$) away.  \citet{Wat2011} mapped the distribution of neutral hydrogen gas in NGC~3906 and found a very circular HI disk similar to that seen at the other wavelengths. There was no evidence of tidal tails or streams which suggests that the off-centered bar in NGC~3906 is likely not from any on-going interaction of the galaxy with any nearby galaxies.

In this paper, we investigate the origin of the off-centered bar in NGC~3906 using 3.6 and $4.5\mu$m imaging from the \spitz\ Survey of Stellar Structure in Galaxies (\sfg)\citep{Sheth2010}. These observations allow us to peer through the layers of dust so that we are looking at the dominant older stellar population in the galaxy. The observations and basic properties of the galaxy are described in \S2. More clues to the structure of NGC~3906 are obtained from the photometry of the \spitz\ 3.6 and $4.5\mu$m images. These results are presented in \S3 together with the properties of the stellar bar.

\begin{center}
\begin{table}[h]
  \caption{Properties of NGC~3906. \label{ngc3906_param} }
  \begin{tabular}{@{}lr@{}}
  \tableline
RA (J2000) & 11:49:40.5   $^1$\\
DEC (J2000) & +48:25:33.0 $^1$\\
Morphological Type & SB(r'l)dm  $^2$\\
Distance & 13.5 (Mpc)  $^3$\\
$m_B$ (mag) & 13.5   $^4$\\
$M_B$ (mag) & -17.8 $^5$\\
$D_{25}$  (arcsec) & 112 $^1$\\
Inclination (deg) & $3\pm 2$ $^1$  \\
PA (deg) & 13.5 $^1$ \\
$v_{\rm sys}$ ($\rm km~s^{-1}$) & $959.4\pm0.7$ $^6$ \\
$M_{\rm H\thinspace\footnotesize{I}}$ ($\times 10^8$ \Msun) & 3.76 $^6$ \\
Physical Scale & 1$\arcsec$ = 65 pc
\tablecomments{References: $^1$\citet{Vau1991}; \\
$^2$\citet{Buta2010}; \\
$^3$Distance (in Mpc) calculated from the measured recession velocity from \citet{Vau1991} and corrected for Virgo-centric infall assuming $H_0 = 71\rm~ km~s^{-1}~Mpc^{-1}$; \\
$^4$Apparent $B$-band magnitude corrected for Galactic and internal extinction from \citet{Vau1991}; \\
$^5$Absolute $B$-band magnitude using the apparent magnitude and distance above\\
$^6$Systemic velocity and \HI\ mass measured by \citet{Wat2011}.} 
\end{tabular}
\end{table}
\end{center}

\section{Observations and Archival Data}

The \spitz\ Survey of Stellar Structure in Galaxies (\sfg) is a volume-, magnitude-, and size-limited ($\rm d<40~Mpc$, $\rm |b| > 30^{\circ}$, ${\rm m}_{Bcorr} <15.5$, $ D_{25} >1\rm\arcmin$) imaging survey of over 2300 galaxies using the Infrared Array Camera (IRAC) at 3.6 and 4.5$\mu$m.  The basic properties of the galaxy are listed in Table~\ref{ngc3906_param}. The apparent optical diameter of NGC~3906 is $ D_{25}\sim1.9\rm\arcmin$ and with the  5\arcmin\ field of view of \spitz , the galaxy is mapped well beyond a radial extent of $1.5\times  D_{25}$ within a single pointing.  The \sfg\ survey and data analysis are described in detail in \citet{Sheth2010}. 

\begin{figure*}
   \begin{center} 
\includegraphics[width = 16cm]{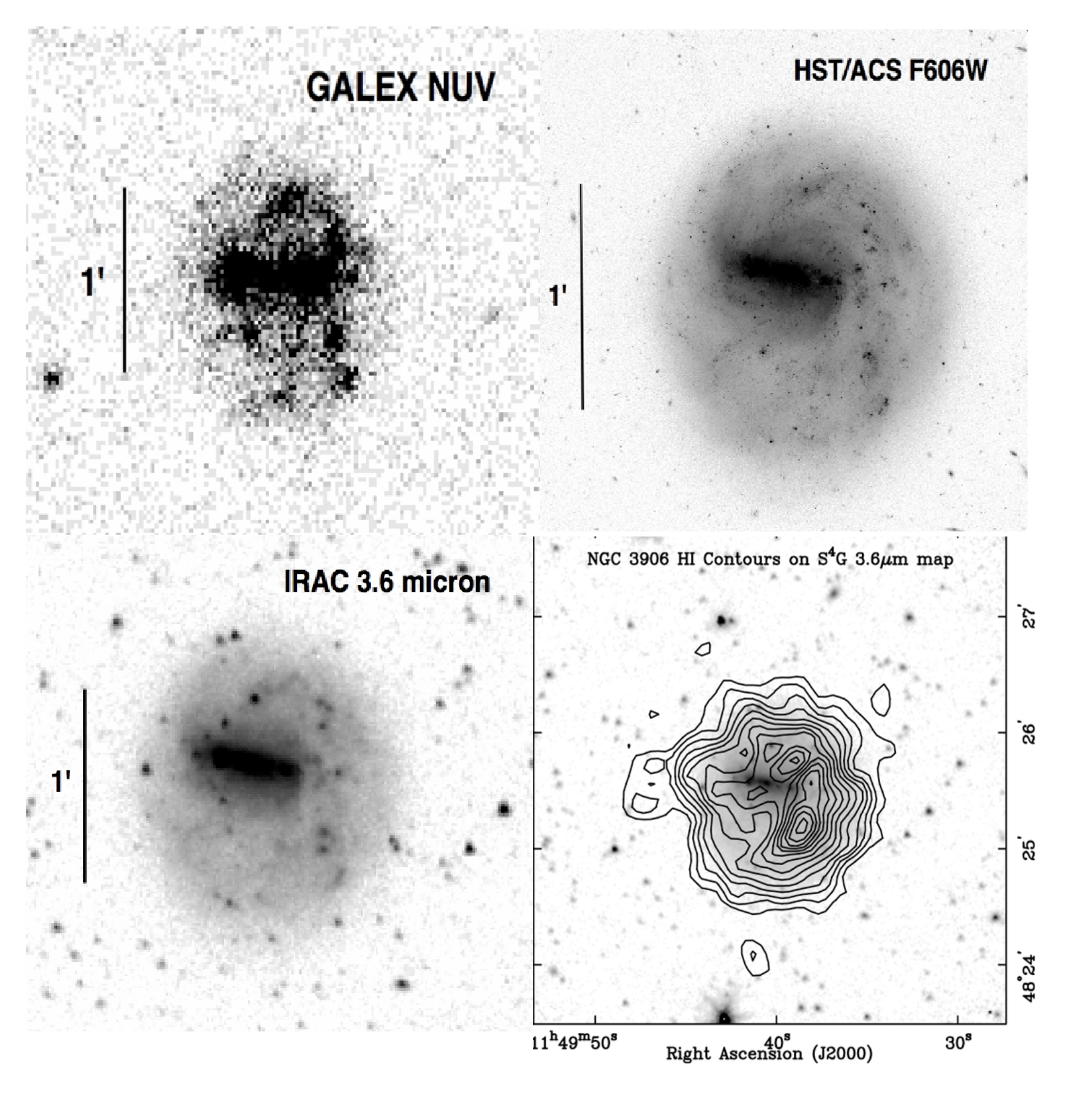}
\caption{Imaging of NGC~3906 at different wavelenths: near-ultraviolet imaging with GALEX (\textit{top left}); at 6060\AA\ with \textit{HST}/ACS using the F606W filter (\textit{top right}); 3.6$\mu$m \sfg\ image with IRAC onboard \spitz\ (\textit{bottom left}); and \HI\ contours are overlayed on the 3.6$\mu$m image (\textit{bottom right}). A scale of 1 arcmin (correspnding to a physical scale of 65~pc) is indicated on the left of each image. North is up and East is left. \label{ngc3906_comparison}}
\end{center}
\end{figure*}

The 3.6$\mu$m \sfg\ image of NGC~3906 is shown in Fig.~\ref{ngc3906_comparison}, alongside images from near-ultraviolet to radio wavelengths (GALEX, HST-ACS (F606W), VLA-HI). The HST image was retrieved from the Multimission Archive at STScI\footnote{http://archive.stsci.edu/index.html .} under the program ID~10829 (PI: Martini, P.). An undisturbed, circular stellar disk is evident at all wavelengths while the spiral features in the galaxy can most clearly be distinguished in the F606W HST image.  The galaxy is face-on  with an inclination of $<$5 degrees as measured in the RC3 and the \sfg\ data.  

\begin{figure}[t]
   \begin{center} 
\includegraphics[width=7.8cm]{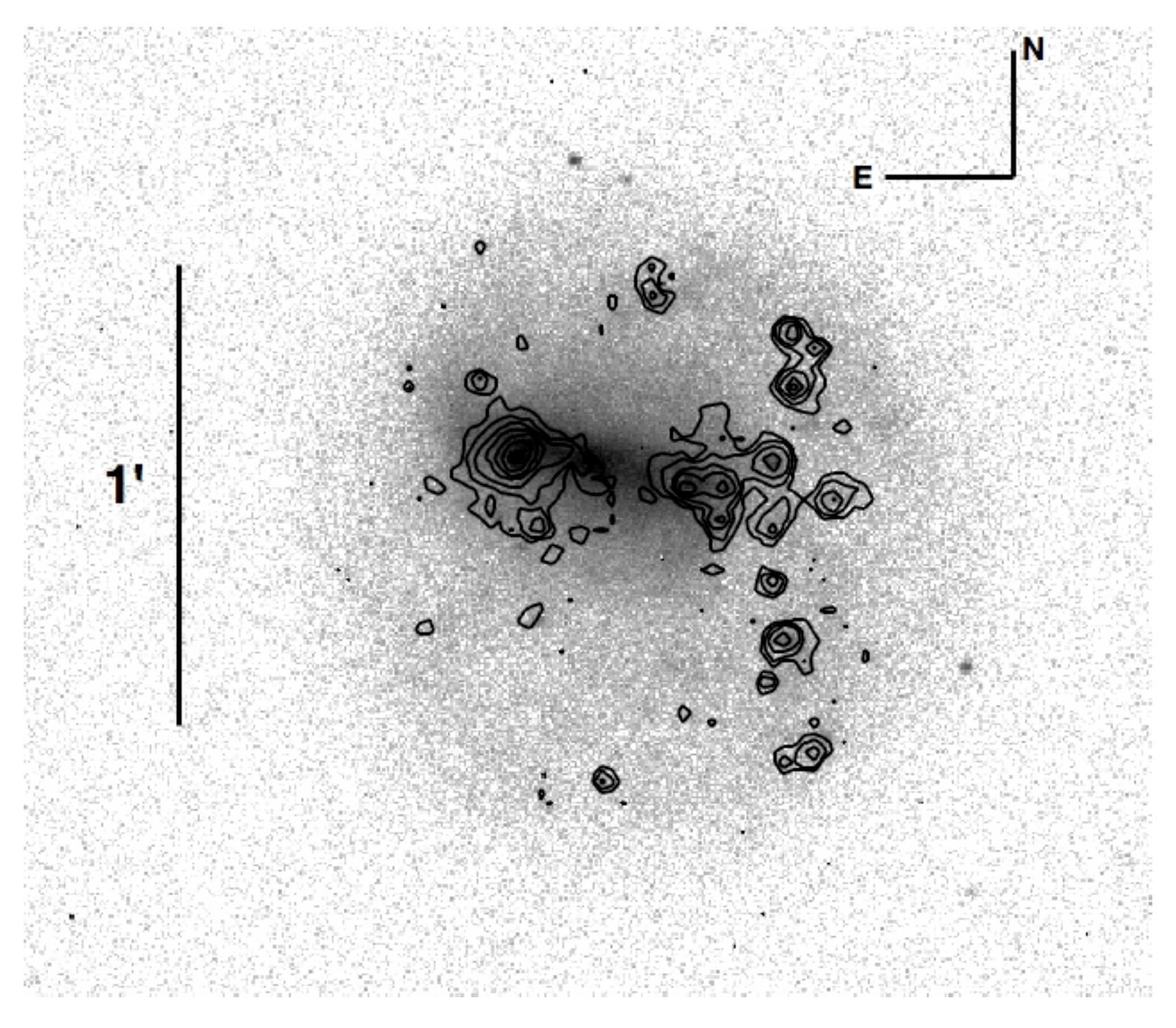}
   \caption{\HA\ contours are over-plotted on the $R$-band image of NGC~3906 to indicate where massive star formation occurs in the galaxy. \label{contours}}
\end{center}
\end{figure}

Narrow-band \HA\ and Johnson $R$-band imaging was obtained by \citet{James2004} using the 1.0m Jacobus Kapteyn Telescope (JKT) at La Palma. The \HA\ contours are over-plotted on the $R$-band image of the galaxy in Fig.~\ref{contours}. These contours together with the near-ultraviolet image from GALEX indicate that massive star formation occurs at the bar ends, a result commonly seen in Magellanic type (e.g. 30 Dor in the LMC) and strongly-barred galaxies \citep[\eg][]{Sheth2000,Sheth2002,James2004}.  Smaller knots of star formation are also observed in the galaxy disk which appear to be associated with the dominant spiral arm. \citet{James2004} have determined a star formation rate of $0.5~\MSUN~\yr$ using the \HA\ imaging of NGC~3906 shown in Fig.~\ref{contours} and assuming a \citet{Kenn1994} star formation rate calibration for disk galaxies. The low star formation rate detected in NGC~3906 is typical of Magellanic spirals, which are believed to be on the verge of exhausting their reservoirs of cold gas. The contours of neutral hydrogen \HI\ gas are overplotted on the 3.6$\mu$m image of NGC~3906 in Fig.~\ref{ngc3906_comparison}. The gaseous disk appears circular and unperturbed, however there is a small extension to the east of the disk which may indicate that NGC~3906 is experiencing an interaction. 

\section{Photometric Characterization from Mid-infrared Imaging}

\subsection{Structure of NGC~3906} \label{structure}

\begin{figure}[t]
	\includegraphics[height = 9cm]{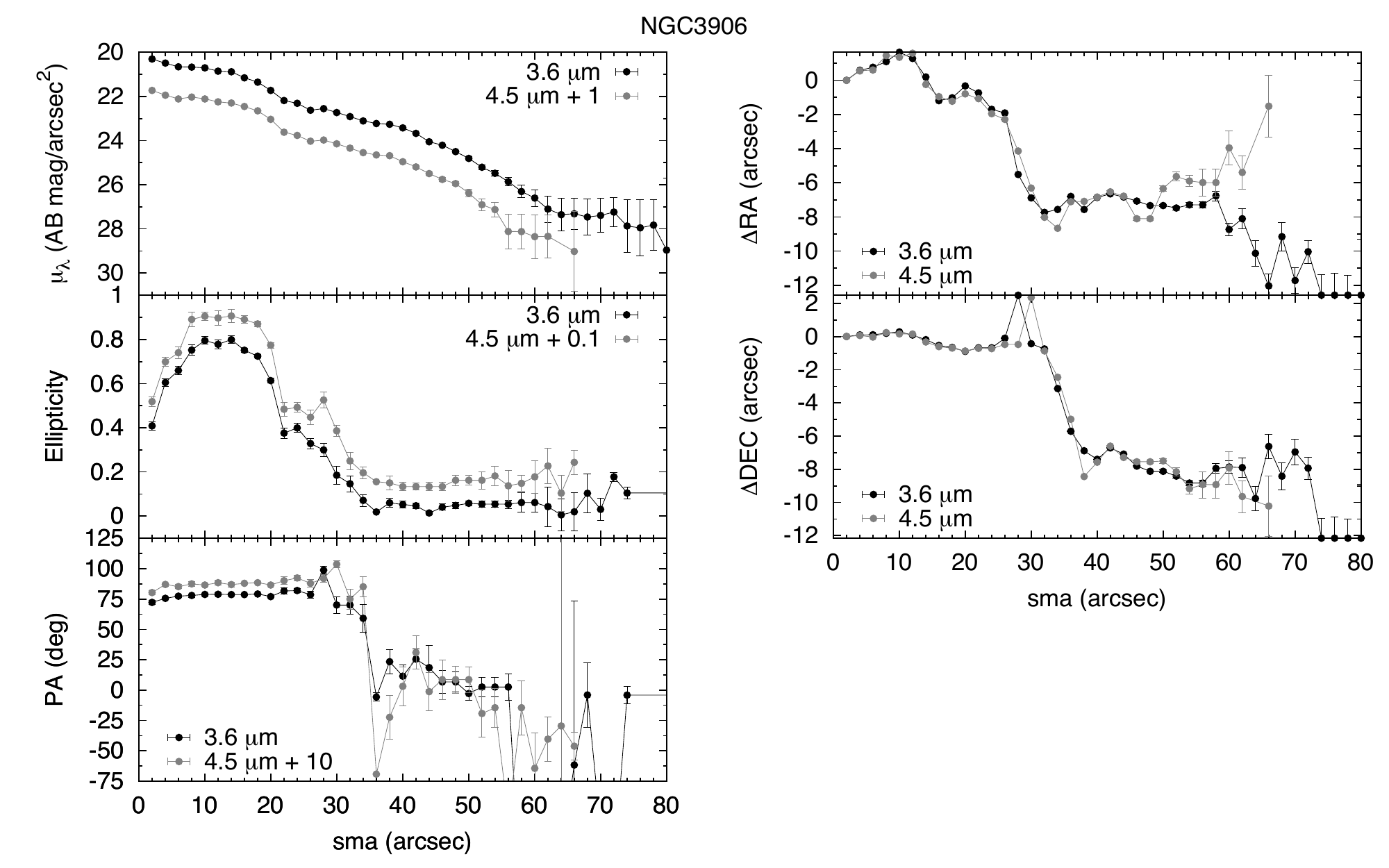} 
	\caption{Radial profiles for NGC~3906 obtained from the surface photometry of the $3.6$ and $4.5\mu$m \sfg\ images. An offset has been applied to the $4.5\mu$m profile for the sake of clarity.\textit{Top Left}: Surface brightness profiles from the $3.6\mu$m (black filled circles) and $4.5\mu$m-bands (grey filled circles). \textit{Middle Left}: Variation of the ellipticity of the best-fitting isophote with the radius. \textit{Bottom Left}: Position Angle (PA) of the best-fitting isophote as a function of radius. \textit{Right}: Change in the centroid of the best fit rllipse in right ascension and declination.  \label{ngc3906_geoprofiles}}
\end{figure}
\begin{figure}[t]
	\includegraphics[height = 8cm]{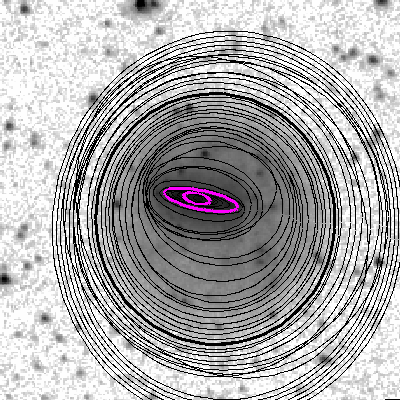} 
	\caption{The elliptical fits at each radius are displayed on the 3.6$\mu$m image of NGC~3906.  The bar isophotes are indicated in purple.  The darker ellipse indicates the $\mu_{3.6\mu m} = 25.5\rm~mag~arcsec^{-2}$ isophote which was used to determine the photometric center of the galaxy disk.  \label{ngc3906_ellipses}}
\end{figure}

We carried out a detailed photometric analysis of NGC~3906 using both the 3.6 and 4.5$\mu$m \sfg\ images to characterize the underlying structure of the galaxy (see details of the method in \citealt{Sheth2010, Munoz15}) . The radial profiles of surface brightness, ellipticity and position angle were determined in both bands using the IRAF\footnote{IRAF is distributed by the National Optical Astronomy Observatory, which is operated by the Association of Universities for Research in Astronomy (AURA) under cooperative agreement with the National Science Foundation.}
 task ELLIPSE and are shown in Fig.~\ref{ngc3906_geoprofiles}. The galaxy center, ellipticity ($\epsilon$) and position angle were allowed to vary when deriving the radial profiles. The elliptical isophotes corresponding to each fit with radius are illustrated in Fig.~\ref{ngc3906_ellipses}. Here it can be seen that the inner isophotes are more elliptical and map the structure of the bar, whereas the outer isophotes trace the circular geometry of the stellar disk.        



The radial profiles shown in Fig.~\ref{ngc3906_geoprofiles} were constructed using a 2 arcsec interval between the successive elliptical fits. The high spatial sampling of these profiles allows us to accurately measure the geometry of the outer stellar disk down to the detection limit of $\mu_{3.6\mu m}(AB)(1\sigma)$= 27~mag~arcsec$^{-2}.$ The surface brightness profile shows that NGC~3906 hosts an exponential disk whereas a more dominant contribution to the light profile from the stellar bar is seen at smaller radii of $r\le20\rm~arcsec$. The circular geometry of the stellar disk is evident for radii $r\ga 40\rm~arcsec$ where the outer isophotes are seen to be nearly circular ($\epsilon\simeq0$). The photometric center of the stellar disk therefore coincides with the geometric center of the  outer isophotes. We have taken the center of the isophote corresponding to $\mu_{3.6\mu m} = 25.5\rm~mag~arcsec^{-2}$ in surface brightness (at a radius of $r = 54\arcsec$) as the center of the stellar disk (see Table~\ref{phot_param}). 

The influence of the stellar bar on the luminosity profile of the galaxy is seen as an increase in the ellipticity for radii $r< 40\rm~arcsec$. The geometrical parameters of the bar were determined from the elliptical fits in the radial range of $r = 7$ to $r = 21~\rm arcsec$. Small variations in the ellipticity ($\Delta\epsilon\la0.06$) are observed in this radial range so that the structure of bar itself is revealed. The geometrical parameters of the bar as measured in this radial range are given in Table~\ref{phot_param}. These represent the mean values in the ellipticity, position angle and central coordinates. The outer radius defines the length of the bar which is 21\arcsec(or $\sim$1.4 kpc). At smaller radii, the isophotal fitting is strongly influenced by star-forming regions inside the bar. 

\begin{figure}[ht!]
 	\includegraphics[width=15cm]{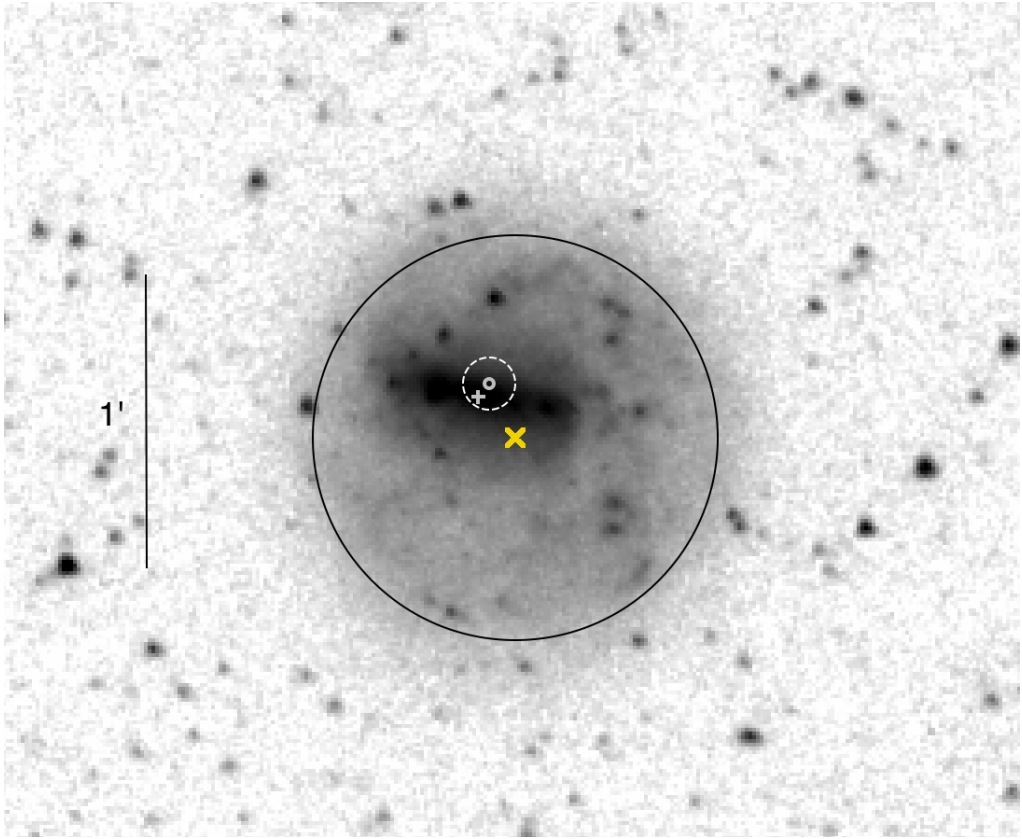}
       	\caption{The location of the photometric centers and dynamical center are overlaid on the $3.6\mu$m image of NGC~3906. The center of the stellar disk (yellow cross) is defined as the center of the $\mu_{3.6\mu m} = 25.5\rm~mag~arcsec^{-2}$ isophote (black solid circle). The dynamical center (open circle) and center of the bar (plus symbol) are indicated in grey. The error associated with the dynamical center is shown by the dashed circle. \label{ngc3906_centers}}
\end{figure}

The photometric center of the stellar disk and that of the bar are shown with a yellow cross and a white plus sign respectively, overlaid on the $3.6\mu\rm m$ image of NGC~3906 in Fig.~\ref{ngc3906_centers}. These positions have a maximum error of 1.2\arcsec\ which accounts for the error in fitting the elliptical isophotes. The displacement of the center of the bar relative to the center of the stellar disk is about $\sim$11 arcsec.  The location of the dynamical center from \HI\ observations \citep{Wat2011} as well as the error associated with this center are shown with a open circle and a dashed error circle respectively  in Fig.~\ref{ngc3906_centers}. The dynamical center overlaps with the photometric center of the bar within the errors of the fit to the HI data ($\Delta$ r = 7$\arcsec$). 
These initial results suggest that the HI gas and underlying stellar component are rotating about a common point which lies close to the photometric center of the bar. 

The \HI\ dynamical center has an angular separation of $\sim14\arcsec$ (or $\sim$ 720 pc) from the photometric center of the stellar disk. This result is in contrast to what is observed for the LMC where the \HI\ dynamical center is offset by $\sim1~\rm kpc$ from the photometric center of the bar \citep{Marel2002}. In addition, \citet{Marel2002} have shown that the bar center in the LMC coincides with the dynamical center of the stellar disk. For the case of the LMC, it has therefore been suggested that the stellar and gaseous components are tracing two different disks. It is also assumed that the dynamical center of the stellar disk is the ``true'' center as the \HI\ distribution is found to be highly disturbed in the gaseous disk of the galaxy due to its on-going tidal interaction with the Milky Way \citep{Kim1998,SS2003}. NGC~3906 does not, on the other hand, show a highly disturbed gaseous disk so that the \HI\ dynamical center likely corresponds with the dynamical center of the stellar disk. \citet{SaloInPrep} have found NGC~3906 to host a very strong bar (bar strength parameter $Q_b = 0.74$ following the method described in \citet{Salo2010} with the force maximum located within the bar region. This provides further evidence that the dynamical center of the stellar disk is found within the bar such that the gaseous and stellar disks have a common center in the bar.  It is therefore very likely that we are observing an offset in the galaxy disk rather than an offset bar in NGC~3906. More clues to the structure of the galaxy will be revealed in the next section when carrying out a two-dimensional decomposition of the \sfg\ images.           



\begin{table}[h]
  \caption{Geometrical parameters of the stellar disk and bar from surface photometry. \label{phot_param} }
  \begin{tabular}{@{}lcccc@{}}
  \hline
  & \multicolumn{2}{c}{Photometric Centers} &    \\
 & RA & DEC & Error \\
 & \multicolumn{2}{c}{(J2000)}  &  (arcsec) \\
 \hline
 Disk & 11 49 39.9 & +48 25 25.2 & 0.06 \\
 Bar  &  11 49 40.6 & +48 25 33.5 & 0.75 \\ 
HI center & 11 49 40.4 & 48 25 36 & 7 \\
 \hline
\end{tabular}
\end{table}

\subsection{Two-dimensional Modeling using GALFIT}

\begin{figure*}[ht!]
    \begin{center} 
    \includegraphics[height = 5.6cm]{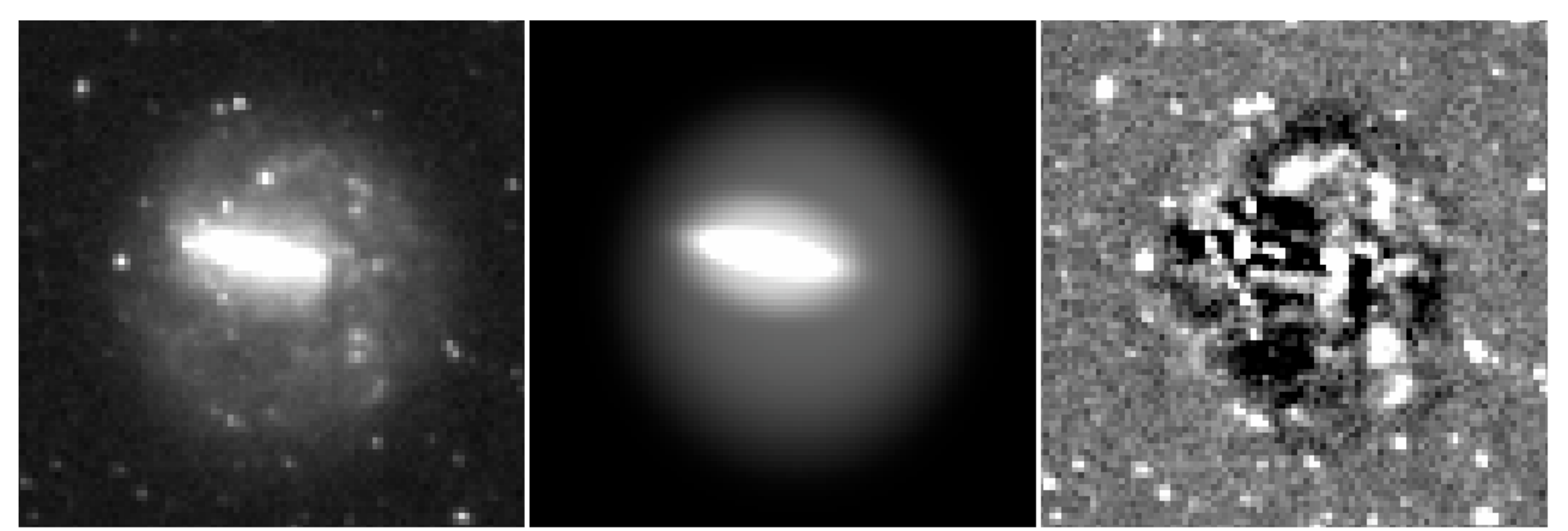}
       	\caption{The three component fit to the light distribution of NGC~3906 using MultiGALFIT. \textit{Left}: The original 3.6$\mu$m image of the galaxy; \textit{Middle}: best S\'ersic profile fits to model the bar, an inner disk and outer disk components. An azimuthal $m=1$ Fourier mode has been added to the disk component; \textit{Right}: residuals from the three component fit of the galaxy.   The residuals are primarily from hot dust around star forming regions that remains a contaminant for 3.6$\mu$m images (see Meidt et al. 2011) \label{ngc3906_galfit}}
	\end{center}
\end{figure*}

A multiple component fitting algorithm was implemented in Python which uses GALFIT\footnote{http://users.obs.carnegiescience.edu/peng/work/galfit/galfit.html} \citep{Peng2002} in modeling the light distribution of the galaxy. The algorithm, called ``MultiGALFIT'' (\textit{private communication}, C.Y.~Peng), allows for up to six components to be fitted without making any assumptions of the number of components needed to model the galaxy. An initial guess for the center of the galaxy is the only parameter required to begin the fitting process. The  multiple component fit was carried out on the binned 3.6$\mu$m image of NGC~3906. Figure~\ref{ngc3906_galfit} shows that NGC~3906 is best modeled (goodness of fit $\chi^2_{\nu} = 0.202$) using a three component fit consisting of three different S\'ersic functions. The parameters for each component fit are listed in Table~\ref{galfit_param}. The innermost S\'ersic function ($r_{\rm eff} = 5.9\arcsec$, $n = 0.41$) models the bar in the galaxy, while the stellar component surrounding the bar, an inner disk, was suitably modeled with a S\'ersic law having an effective radius of $r_{\rm eff} = 7.7\arcsec$ and structural index of $n = 0.23$.  The low sersic indices are typical for late type galaxies as these disks are not centrally concentrated (e.g. \citealt{macarthur2003})
The faint outer disk is displaced from both these components as can be clearly seen in Fig.~\ref{ngc3906_galfit}.

\begin{table}[]
  \caption{Derived parameters for stellar component fits using multiGALFIT. \label{galfit_param} }
  \begin{tabular}{@{}lccc@{}}
  \hline
  &  \multicolumn{3}{c}{S\'ersic Parameters} \\
Stellar  & $r_{\rm eff}$ &  & $M_{3.6\mu\rm m}$ \\
Component & (arcsec) & $n$ & (\Msun) \\  
 \hline
 Bar  & 5.9 & 0.41 & $2.99^{~\pm 0.073}$ $\times 10^8$ \\
 Inner disk   & 7.7 & 0.23 & $5.38^{~\pm 0.071}$ $\times 10^8$ \\ 
 Outer disk  & 14.4  & 0.23 & $1.90^{~\pm 0.069}$ $\times 10^{9}$ \\ 
 \hline
\end{tabular}
\end{table}

The extended stellar disk of the galaxy is best modeled using a S\'ersic function with an effective radius of $r_{\rm eff} = 14.4\arcsec$ and index $n = 0.23$. Perturbations were introduced in to the model in the form of the azimuthal $m=1$ Fourier mode to obtain a measure of the asymmetry of the stellar disk. In face-on systems (inclinations of $i\la25\deg$) such as NGC~3906, the amplitude of the Fourier $m=1$ mode ($A_1$) is an excellent indicator of lopsidedness of the disk \citep{Rix1995}. The projection effects are minimized in these low inclination galaxies so that the apparent structure of the disk is revealed. The 3.6$\mu$m image of NGC~3906 allows for the measurement of the Fourier $A_1$ amplitude in the extended stellar disk out to at least three times the effective radius of the disk. An amplitude of $A_1 = 0.18\pm0.01$ was measured for the disk component using MultiGALFIT, consistent with the \citet{Zar2013} measurement in the outer disk as shown in their Figure 6.  \citet{Bou2005}  measured a mean Fourier $m=1$ amplitude of 0.1 for a sample of 149 spiral galaxies in the near-infrared with values above this threshold indicating lopsidedness. \citet{Zar2013} have done the same for 167 face-on galaxies from \sfg as noted earlier.  Compared to other disks, NGC 3906 shows a modest amount of lopsidedness in its outer disk but as \citet{Zar2013} note the lopsidedness is typically uncorrelated with the presence of the bar.

\begin{figure}[ht]
    \begin{center} 
    \includegraphics[height = 8cm]{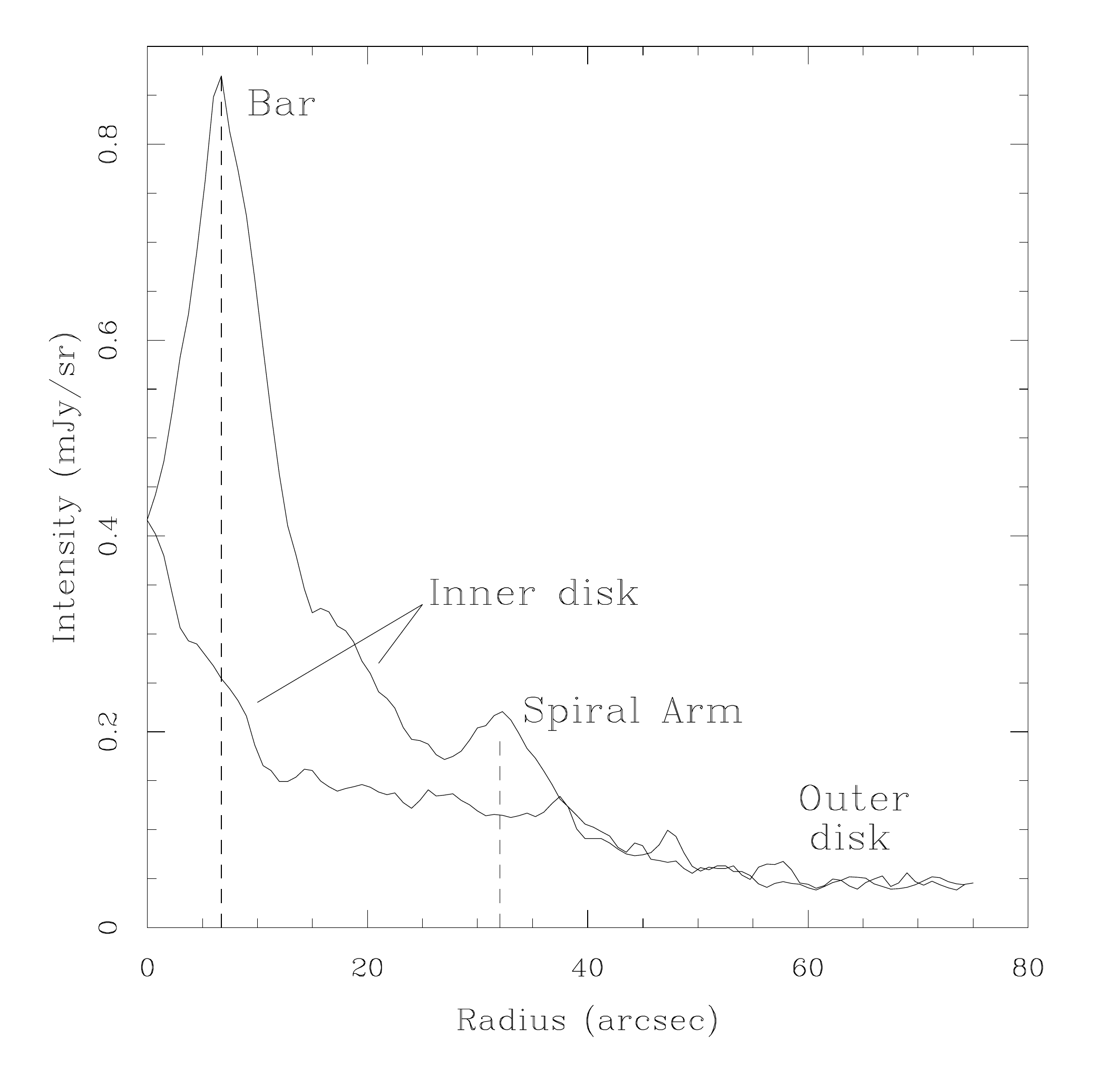}
       	\caption{A one-dimensional (1D) profile cut which runs perpendicular to the bar and through the photometric center of the galaxy. The radius $r=0\arcsec$ corresponds to the photometric center of NGC~3906. The 1D profiles on either side of the photometric center (top one is the northern side) are shown to emphasize the dramatically different light distribution due to the offset bar.    \label{ngc3906_cut}}
	
	\end{center}
\end{figure}

By making a cut perpendicular to the bar and through the photometric center of the galaxy (given in Table~\ref{phot_param}) the symmetry of the disk could be further investigated. The opposite sides of the one-dimensional cut about the photometric center are plotted in Fig.~\ref{ngc3906_cut} where the strongest peak in the intensity distribution corresponds to the bar itself on the northern side of the galaxy. The intensity peak at a radius of $r\sim32\arcsec$ coincides with the spiral arm feature in the northern part of the stellar disk. 

\begin{figure}[ht]
    \begin{center} 
    \includegraphics[height = 6cm]{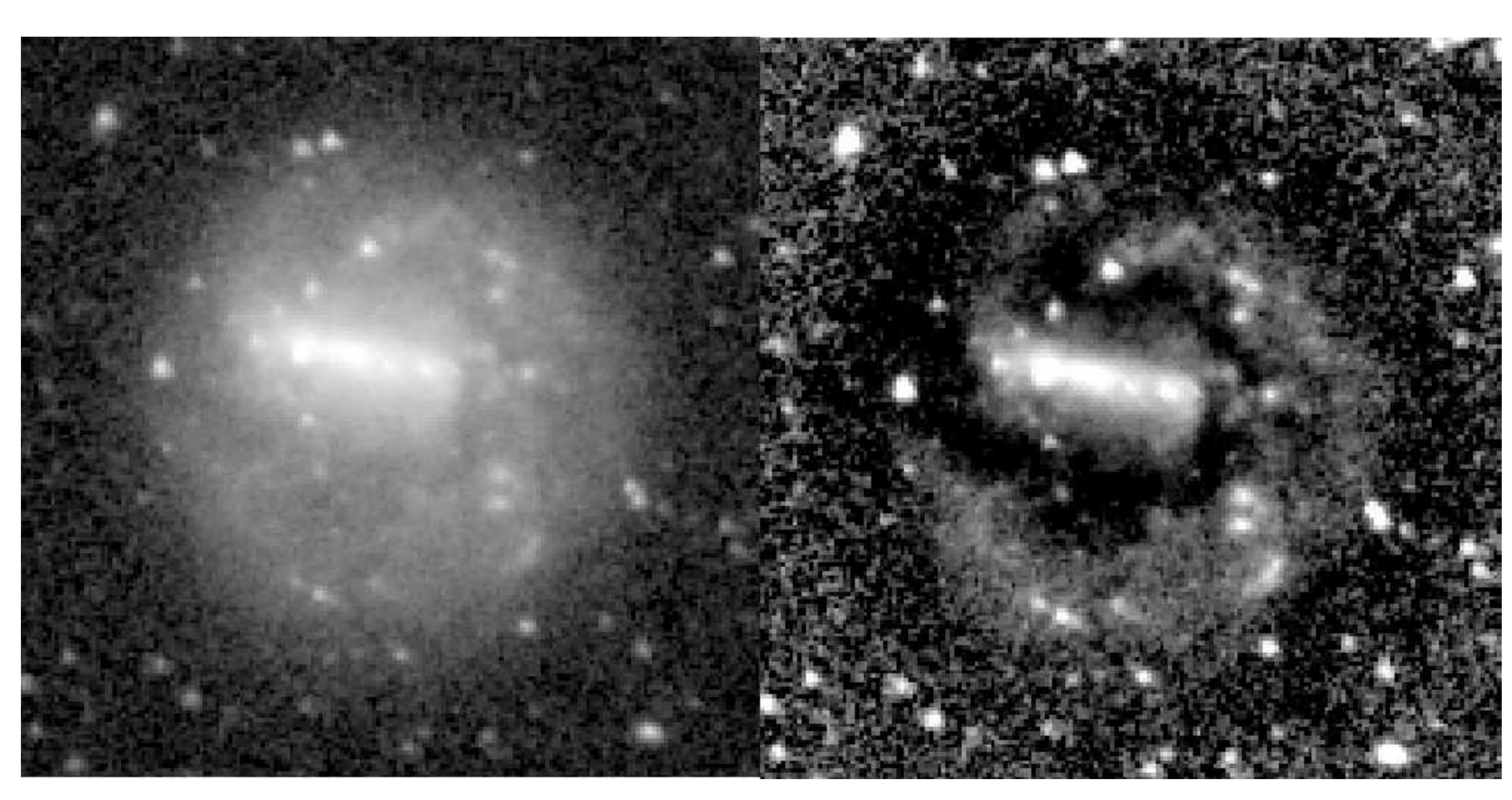}
       	\caption{Fine structure in NGC~3906 using unsharp masking. The original 3.6$\mu$m image is shown on the left while the residual image is shown on the right. \label{ngc3906_unsharp}}
	\end{center}
\end{figure}

The technique of unsharp masking \citep{Mal1977} was applied to the $3.6\mu$m \sfg\ image to reveal  underlying structure in the stellar disk. The features are detected by dividing the original galaxy image by a smoothed image of the galaxy. The unsharp mask for NGC~3906 was produced by smoothing the $3.6\mu$m galaxy image with an elliptical Gaussian of kernel size $\sigma_{kernel} = 15$ pixels.   Figure~\ref{ngc3906_unsharp} shows the residual image of NGC~3906 after carrying out the unsharp masking. The stellar bar and inner disk can be clearly seen in the residual image. A single bright dominant spiral arm is seen to originate from the east side of the bar and winds in a clockwise direction in the disk. Fainter arms can be seen extending from the bar edges from both the east and west regions of the bar. The fainter arms eventually merge with the dominant spiral form. The asymmetric spiral structure is seen as the unsharp mask makes prominent the single, bright spiral arm (as opposed to symmetric grand design spiral arms).  This is characteristic in late-type spirals and Magellanic-type galaxies \citep{Vau1970}.  However it is unclear what phenomenon gives rise to these types of spiral features typical in late-type Magellanic systems.  



\subsection{Mass Estimates of various components}

The mass of the individual stellar components were calculated from the total fluxes measured from MultiGALFIT. The luminosity of each stellar component was determined using the zero point flux of 280.9~Jy for IRAC channel 1 \citep{Reach2005} and an absolute magnitude of $M_{\odot}^{3.6\mu\rm m} = 3.24~\rm mag$ for the Sun in the IRAC $3.6\mu$m-band \citep{Oh2008}. Both the LMC and NGC~3906 are late-type disk galaxies having similar morphology and stellar populations. The mass-to-light ($M/L_{3.6}$) ratio obtained for the LMC was therefore adopted when transforming luminosity in to mass in NGC~3906. The $M/L_{3.6}$ of 0.5 for the LMC from \citet{Eskew2012} was used to derive the mass for the stellar components listed in Table~\ref{galfit_param}. 
These results show that NGC~3906 has a massive disk in which the bar and inner disk contribute to less than $50\%$ to the mass of the extended stellar disk. Adding the mass estimates of the three stellar components and the \HI\ mass in Table~\ref{ngc3906_param} gives a total baryonic mass of $\sim3 \times 10^{9}\MSUN$ for NGC~3906 which is comparable to the mass estimates obtained for the LMC and other barred Magellanic spirals in \citet{Wil2004}.  



\section{Discussion}

Tidal interactions remain as the most likely explanation for the offset between the kinematic and photometric centers of NGC 3906.  New numerical simulations from our team are finding that an interaction with a body at least 1/10th the mass of the main galaxy is needed to create the observed offset (Pardy et al. 2015, in preparation),  consistent with previous studies \citep{Ath1996,Ath1997,Ber2003}.  However the main problem for NGC 3906 is that observations in a wide range of wavelengths rule out a nearby companions.  This is also the case for many other Magellanics with offset bars \citep{Wil2004}.  An interaction with a dark companion such as a dark matter sub-halos could explain the offset as suggested by  \citet{Bek2009} but the constraints on the type of collision required may be that the companion galaxy has to pass close to the center of the main galaxy (e.g.  \citep{Bes2012}).  We are currently studying the precise range of interaction / merger parameters that can lead to an offset bar to constrain the nature of the dark matter substructure.  We are also investigating how long the observed offset can last in different types of collisions to better constrain the evolution of this system.

NGC 3906 is also a lopsided disk with the strongest asymmetry in the inner parts evidently from the offset between the bar and the disk.  \citep{Zar2013} have already shown that the general class of asymmetries (m=1 Fourier modes) is too common to exist from purely interaction events.  The asymmetry  must also be long lived since it is commonly observed.  Noordemeer et al (2001) and Levine \& Sparke (1998) have argued that long-lived lopsidedness may occur if the stellar disk is off-center from a galaxy's dark matter halo. However, they do not provide any reason or mechanism which would cause the disk to be displaced in the first place.   Secular and internal processes such as gas accretion can also lead to lopsidedness in the disk \citep{Bou2005}.  


\section{Conclusions}

Images from \sfg\ of NGC~3906 reveal a strikingly odd offset between the stellar bar and the underlying stellar disk.  Analysis of the \sfg\ data and VLA HI data reveal that, unlike the LMC, the  HI kinematic center is coincident with the bar, but is offset from the photometric center of the stellar disk by $\sim$720 pc.  Investigation of the bar structure using ellipse fits and a full 2-D decomposition using MultiGALFIT using a bar, inner and outer disk components  reveals a strong 1.4 kpc bar (Q$_{b}$ = 0.74)  containing $\sim$10\% of the total baryonic mass (M$_{t} \sim$ 3 $\times$ 10$^9$ \Msun) of the galaxy. Residual images of the galaxy also show the presence of asymmetric spiral structure.   Given the lack of any nearby companions, a likely  explanation for the offset may be an interaction with a dark matter sub halo.  We are carrying out new simulations to create the observed offsets and find that  a companion with at least 1/10th the mass of the disk is needed.  We are currently studying the lifetime of the offset to better constrain the likely evolution of this system.  Other observations have shown that many Magellanic systems with offset bars are also similarly isolated and therefore may offer a unique probes for the structure of the dark matter halo if collisions with dark matter sub halos are responsible for offset bars.  

\section{Acknowledgements}

We thank the anonymous referee for comments and feedback that greatly improved this paper.  K. Sheth, J.C. Munoz-Mateos, T. Kim and B. de Swardt gratefully acknowledge support from the National Radio Astronomy Observatory which is a facility of the National Science Foundation operated under cooperative agreement by Associated Universities, Inc.  BdS would also like to acknowledge the South African National Research Foundation (NRF) for financial support provided for the research presented in this paper.  ED gratefully acknowledges the support of the Alfred P. Sloan Foundation and the Aspen Center for Physics for their hospitality and support under grant No. PHYS-1066293.   ED and SP acknowledge support provided by the University of Wisconsin-Madison Office of the Vice Chancellor for Research  and Graduate Education with funding from the Wisconsin Alumni Research Foundation and  NSF Grant No. AST-1211258 and ATP-NASA Grant No. NNX14AP53G.  EA, AB, AGdP, JHK, EL,  ES and HS acknowledge  financial  support  from  the People  Programme (Marie Curie Actions) of the European Union’s  Seventh Framework Programme FP7/2007-2013/ under REA grant agreement number PITN-GA-2011-289313 to the DAGAL  network. EA and AB also ackowledge financial support from the CNES  (Centre National d’Etudes Spatiales - France) and from the “Programme  National de Cosmologie et Galaxies” (PNCG) of CNRS/INSU, France.  JHK acknowledges financial support from the Spanish Ministry of Economy and Competitiveness (MINECO) under grant number AYA2013-41243-P.

\end{document}